# Long-term mutual phase locking of picosecond pulse pairs generated by a semiconductor nanowire laser

B. Mayer[1], A. Regler[1,2], S. Sterzl[1], T. Stettner[1], G. Koblmüller[1], M. Kaniber[1], B. Lingnau[3], K. Lüdge[3] and J. J. Finley[14]

[1]*Walter Schottky Institut and Physik Department, Technische Universität München, Am Coulombwall 4, Garching 85748.*
[2]*Institute for Advanced Study, Technische Universität München, Lichtenbergstrasse 2a, Garching, Germany, 85748.*
[3]*Institute of Theoretical Physics, Technische Universität Berlin, Hardenbergstraße 36, 10623 Berlin.*
[4]*Nanosystems Initiative München, Schellingstraße 4, München, Germany, 80799.*

**The ability to generate phase-stabilised trains of ultrafast laser pulses by mode-locking underpins photonics research in fields such as precision metrology** [1] **and spectroscopy** [2] [3] [4] [5]**. However, the complexity of conventional mode-locked laser systems, combined with the need for a mechanism to induce active or passive phase locking between resonator modes, has hindered their realisation at the nanoscale** [6]**. Here, we demonstrate that GaAs-AlGaAs nanowire lasers** [7] [8] **are capable of emitting pairs of phase-locked picosecond laser pulses when subject to *non-resonant* pulsed optical excitation with a repetition frequency up to ~200GHz** [9]**. By probing the two-pulse interference that emerges within the homogeneously broadened laser emission** [10] [11]**, we show that the optical phase is preserved over timescales extending beyond ~30ps, much longer than the emitted laser pulse duration (~2ps). Simulations performed by solving the optical Bloch equations produce good quantitative agreement with experiments, revealing how the phase information is stored in the gain medium close to transparency. Our results open the way to applications such as on-chip, ultra-sensitive Ramsey comb spectroscopy** [12] [13] [1]**.**

Wavelength scale coherent optical sources are vital for a wide range of applications in nanophotonics ranging from metrology [14] and sensing [15], to nonlinear frequency generation [16] and optical switching [17]. In these respects, semiconductor nanowires (NWs) are of particular interest since they represent the ultimate limit of downscaling for photonic lasers with dielectric resonators [12]. By virtue of their unique one-dimensional geometry NW-lasers combine ultra-high modal gain, support low-loss guided modes and facilitate low threshold lasing tuneable across the UV, visible and near infra-red spectra regions [11] [7] [8] [18]. Recently, optically pumped NW lasers have been demonstrated at room temperature and they can now be site-selectively integrated onto silicon substrates [9] [19]. While the fundamental carrier relaxation and gain dynamics of NW-lasers have been very recently explored [10] [11], the coherent dynamics have hitherto received comparatively little attention. Here, we directly probe the mutual coherence between subsequently emitted ultrafast pulses from GaAs-AlGaAs core-shell NW-lasers. Two-pulse interference that emerges within the homogeneously broadened NW laser line recorded in time integrated measurements show that the optical phase is preserved over timescales that are $\geq 10 \times$ longer than the emitted pulse duration. The mechanism responsible for the mutual phase locking is shown to be linked to coherent dynamics that occur in the *post-lasing* regime when the system remains close to transparency over timescales approaching the spontaneous emission lifetime. Numerical simulations performed by solving the optical Bloch equations produce good quantitative agreement with experiments supporting this conclusion.

The structures investigated are GaAs-AlGaAs core-shell nanowires (NWs) grown using solid-source molecular beam epitaxy on silicon substrates [8]. After growth, individual NWs were mechanically removed from the growth substrate and dispersed onto glass whereupon the lasing behaviour of individual NWs could be explored when subject to optical pumping. In experiments performed using single excitation pulses, a clear transition from spontaneous emission to single mode lasing is observed upon increasing the excitation level, with a typical threshold pulse energy of $P_{th} \sim 9$ pJ per pulse for lasing (see supplemental fig S1). To study the ultrafast emission and coherence properties of the NW-lasers, we performed time resolved pump-probe spectroscopy with non-resonant excitation, as a function of the energy of the pump ($P_{pump}$) and probe ($P_{probe}$) pulses, respectively. Hereby, the NW-lasers were excited using ~250fs duration laser pulses at a repetition frequency of 80MHz tuned to selectively excite the active GaAs-core region of the NW *non-resonantly* at $\hbar\omega_{exc}$=1.59eV [7] [8] [9]. The temporal separation between pump and probe was precisely tuneable over the range $\Delta t = \pm 100$ ps using an optical delay line that provides a relative precision better than $\sim 10$ fs and the experimental observable was the time integrated emission spectrum averaged over $\geq 10^6$ pump-probe excitation pulse pairs.

The leftmost panels of fig. 1 show typical pump-probe data recorded as a function of the time delay between pump and probe ($\Delta t$) for $P_{pump}$ close to $\sim 3 \times P_{th}$ and three different values of $P_{probe}$; below threshold (upper panel), close to threshold (middle panel) and above threshold (lower panel). Pronounced interference fringes are observed in the time-integrated spectra recorded for all three combinations of $P_{pump}$ and $P_{probe}$ and persist even for $\Delta t \geq 40 ps$. To check that the observed fringes do indeed arise from the interference between two successively emitted laser pulses, the fringe separation in the frequency domain $\Delta f$ is plotted in fig 2 as a function of $\Delta t$. As can be seen, $\Delta f$ decreases *inversely* with $\Delta t$ as expected, indicative of the interference between subsequently emitted NW laser pulses generated by the pump and probe pulses [20]. The data indicate that extremely high maximum repetition rates $\Delta f$ >200GHz are possible, corresponding to emitted pulse durations $t_{pulse} < 5ps$. Two-pulse interference was recently reported for GaN, CdS, ZnO NW-lasers in operation regimes corresponding to the *direct* temporal overlap between the emitted laser pulses [10] [11]. In contrast, in Fig 1 we note that interference is *still observed* in the time integrated spectrum over timescales more than one order of magnitude *longer* than the duration of the emitted laser pulses themselves. This surprising observation clearly indicates that coherence is preserved in the NW-laser over long timescales after lasing has ceased.

Our experimental observations are in excellent agreement with the predictions of a numerical model of the pulsed, driven laser system obtained in the framework of the well-known semiconductor Bloch equations for lasers [21] [22] with microscopically motivated carrier dynamics [21]. A stochastic spontaneous emission noise source was introduced to simulate the effects of dephasing [21] [22]. The Bloch equations are extended by an additional equation describing the time dependent carrier density in the reservoir at 1.59eV pumped in our experiment and the resulting incoherent scattering processes into the lasing state at 1.51eV [23] (see Methods). The rightmost panel of fig 1 shows selected results of our numerical simulations. The best quantitative agreement with our experimental observations was obtained for a photon lifetime in the resonator of ~1ps, obtained from the measured Q-factor of the resonator modes, carrier lifetimes in the reservoir of ~10ps and a spontaneous emission lifetime of the lasing state equal to ~0.6ns. We note that the simulation reproduces most of the principle features observed in the experiment, including the two-pulse interference and the time dependent redshift caused by a small detuning between lasing transition and cavity mode, measured to be 1.7meV (see Supplementary). The good agreement between experiment and simulations confirm the interpretation of the observed features presented in fig 1(a) as being due to 2-pulse interference in the time integrated spectral response of the NW-lasers studied.

We continue to explore the pump-probe dynamics of the NW-lasers in the time domain by computing the discrete Fourier transform of the experimental spectra presented in Fig. 1. The results of this procedure are presented in Fig. 3 (left panels) together with the corresponding numerical simulations (right panels). The figure compares data recorded for a situation with $P_{pump}$ in the lasing regime and $P_{probe} \ll P_{th}$ in the spontaneous emission (SE) regime, labelled by L-SE (fig 3a), and a situation where *both* $P_{pump}$ and $P_{probe}$ are in the lasing regime labelled L-L (fig 3b). The colour plots reveal that two, temporally distinct NW-laser pulses are always generated, provided that $P_{Pump}$ is chosen to be above threshold (e.g. $P_{pump}/P_{th} > 2$) and the pump pulse arrives *first* to create gain in the NW. For situations when the weaker probe pulse arrives before the stronger pump then only one laser pulse is observed ($\Delta t > 0$ – fig 3a). We note that in the L-SE configuration, the interference observed in the time integrated spectrum is indicative that the residual gain in the NW is sufficient for the weak probe pulse to re-establish lasing and that the two pulses are mutually phase coherent despite exhibiting no temporal overlap at the detector. When both excitation pulses are in the lasing regime (L-L – fig 3b) a clear symmetry is observed around the point of coincidence ($\Delta t = 0$ps) as two pronounced coherent NW laser pulses well separated by a delay >30ps are observed for positive and negative $\Delta t$, labelled pulse-1 and pulse-2 on fig 3. The duration of the first NW laser pulse represented by the white line at the bottom of the panels is measured to be $t_{pulse} = 0.9 \pm 0.1 ps$ whereas the second pulse from the probe excitation has a duration ranging from $t_{pulse} = 3.7 ps$ ($\Delta t = 15$ps) to $1.6 ps$ ($\Delta t = 30$ps). Therefore, the analysis in the time domain is in excellent agreement with the results obtained from the analysis of the two-pulse interference in the frequency domain where similar pulse durations of $t_{pulse} < 5 ps$ corresponding to a ~200GHz fringe separation are expected.

We now turn our attention to the weak periodic oscillation clearly seen in the L-L data presented in fig 3b. These periodic oscillations are attributed to Rabi oscillations that occur during the pulse emission and in the few picoseconds after lasing has ceased. This can be seen in fig. 4(a) that shows the computed occupation of the lasing state as a function of time (lower panel) together with the theoretical (green line) and experimental (black line) temporal evolution of the photon field in the NW-laser (upper panel). The results show that the lasing state is almost fully inverted (occupation>0.8) before the emission of a NW laser pulse, marked by the red region on the figure. The initial NW laser pulse is formed by Rabi oscillations (blue region) and a persistent occupation close to transparency (~0.5) extending over timescales towards the spontaneous emission lifetime. Therefore, the initial NW laser pulse can be considered to be the first beat of a strongly damped Rabi oscillation that prevails for several tens of picoseconds. Figure 4(b) shows the computed electric fields of the first and second emitted pulses in the complex plane using the same parameters that were used to obtain good agreement with experiments in figs 1(b) and 3(b). Clearly, the phases of the two subsequently emitted NW laser pulses overlap perfectly even after $\Delta t$ =20ps (blue lines) and increase only slowly from $\Delta t = 25 ps$ (green lines) to $\Delta t = 30 ps$ (red lines), indicative of a gradual loss of the mutual coherence stored in the coupled electron-photon system after lasing has ceased. We continue to provide a consistent explanation of the phase information transfer between subsequent NW laser pulses. As depicted on fig. 4, after the initial NW laser pulse emission the population of the lasing state remains very close to transparency (~0.5) since stimulated emission and absorption processes occur over similar timescales. As such, the comparatively slow spontaneous emission depopulates the lasing state over much longer timescales (~1ns) and the NW gain medium remains close to transparency for several tens of ps. Consequently, after emission of the first pulse the residual coherence is preserved in the coupled electron-photon system within the quasi transparent NW medium as continuous absorption and stimulated emission is accompanied by collective Rabi oscillations. Hereby, the spontaneously generated phase of the first pulse emitted remains stored within the NW-laser over timescales such for which the intensity remains larger than the noise level. If the NW-laser is subject to a second excitation pulse, relatively little power is needed to lift the population above the lasing threshold and the subsequently emitted pulse recovers the phase of the first pulse emitted.

In summary, we have demonstrated how coherent phase information can be transferred between two subsequently emitted NW laser pulses by virtue of coherent Rabi oscillations that occur during the post-lasing transparency regime. The consequent observation of interference fringes reveals the potential for high repetition rates >200GHz whilst preserving mutual coherence between the emitted laser pulses. The discrete Fourier transform of the spectra obtained from ultrafast pump probe experiments exhibits coherent information storage >30ps and pulse durations <2ps. The ultrafast pulse emission of the NW lasers is followed by strongly damped Rabi oscillations that result from the coherent light matter interactions in the NW cavity. The experimental results are in good quantitative agreement with theoretical predictions from a microscopically motivated semiconductor Bloch equation model with a realistic noise source that calculates the phase of the photon field and the polarisation of the lasing state population in the NW. The theoretical model is capable of fully describing the coherent phase information transfer between two subsequent NW laser pulses and the generation of Ramsey combs found in the experiments. The results present a novel mechanism for optical information storage and frequency comb generation at ultrafast timescales at the nanoscale.

**Methods:**

**Sample preparation:** Details regarding the growth, physical and optical properties of the NW-lasers investigated can be found in reference [8].

**Numerical modelling:** Our model is adapted from the well-known Bloch equations for lasers [24] [25] with an additional equation describing the carrier density in the reservoir $N_2$ and the respective incoherent scattering processes from the reservoir to the lasing state [23] (see Methods). We emphasize that phase coherence is *not* expected to be transferred from the excitation source to the NW laser output, since the relaxation from the excited reservoir states at $\hbar\omega_{exc}$=1.59eV to the lasing state at $\hbar\omega_{em}$=1.51eV is incoherent. The system of equations that are solved read:

$$\dot{p} = -i\Delta\omega\, p - \frac{p}{T_P} - iE\frac{\mu_0}{2\hbar}(2\rho_1 - 1)$$

$$\dot{E} = -ig_m p - \kappa E + D\xi$$

$$\dot{\rho}_1 = R(\rho_0 - \rho_1) - \frac{\rho_1}{T_1} + Im\left(pE^*\frac{\mu_0}{\hbar}\right)$$

$$\dot{N}_2 = J - R(\rho_0 - \rho_1) - \frac{N_2}{T_2}$$

Here, the complex electric field amplitude $E$ is described in the rotating frame of the optical frequency of $\hbar\omega \sim 1.51 eV$ (according to the experimental data), $p$ is the polarization and $\rho_1$ the occupation probability of the electrons inside the lasing transition. The number of carriers excited within this lasing level is given by $\rho_1 N^e$ with the total carrier density $N^e$ that can be accommodated by the lasing state of the nanowire. The dominating timescales are the photon lifetime $\tau_{ph}=(2\kappa)^{-1}=1ps$, the polarization lifetime $T_p=5ps$, and the electron lifetimes of reservoir and lasing level, $T_2$ and $T_1$, respectively. Within the reservoir we assume fast phonon scattering $T_2=10ps$ while within the lasing level the electrons are lost via spontaneous emission, i.e., $T_1=0.6ns$. The relaxation between the two electronic levels is implemented via a relaxation rate approximation, with a relaxation rate of $R^{-1}=2.5ps$ to reach a quasi-equilibrium occupation of $\rho_0$ (given by a Fermi distribution). The spontaneous emission factor ($\beta$) was chosen to be $\beta=0.01$. For the numerical integration the noise is implemented as a stochastic Gaussian white noise source $\xi$ with noise strength $D = \sqrt{\frac{\beta\rho_1 N^e}{T_1}\frac{\omega}{2\epsilon_{bg}\epsilon_0}} = \sqrt{\frac{\beta\rho_1}{T_1}g_m\frac{\hbar}{\mu_0}}$ [22] [21]. The parameter $g_m$ is the coupling constant that determines the gain of the light matter interaction. It depends on the density of available electrons $N^e$ and on the dipole moment $\mu_0$=0.16 e nm of the transition (both have been chosen to yield the gain observed in experiment). The carriers are injected into the reservoir with a pulsed pump intensity $J = J_{pump}e^{-\left(\frac{t-t_0}{\Delta_P}\right)^2} + J_{probe}e^{-\left(\frac{t-t_1}{\Delta_P}\right)^2}$ at times $t_0$ and $t_1$ with a pump-pulse-width of $\Delta_P$=100fs according to the experimental setup.

The value chosen for $g_m$ yields a rate equation gain (modal gain) of 280 cm$^{-1}$ for a detuning of $\Delta\omega$=1THz. To produce the data presented in the main manuscript carriers are first injected into a reservoir and subsequently relax to the optically active level. The input parameters for gain and lifetimes (photons, electrons) where chosen according to experimental details: For example, for the data presented in figs 1, 3 and 4 we chose a photon lifetime $\tau_p$=1ps, a lifetime of electrons within the lasing level $T_1$=0.6ns, an electron lifetime inside the reservoir $T_2$=10ps and a relaxation rate between reservoir and lasing level of $R^{-1}$=2.5ps. The polarization lifetime of $T_p$=5ps was adjusted to reproduce the experimental results (longer/shorter $T_p$ lead to more/less visible Rabi-oscillations).

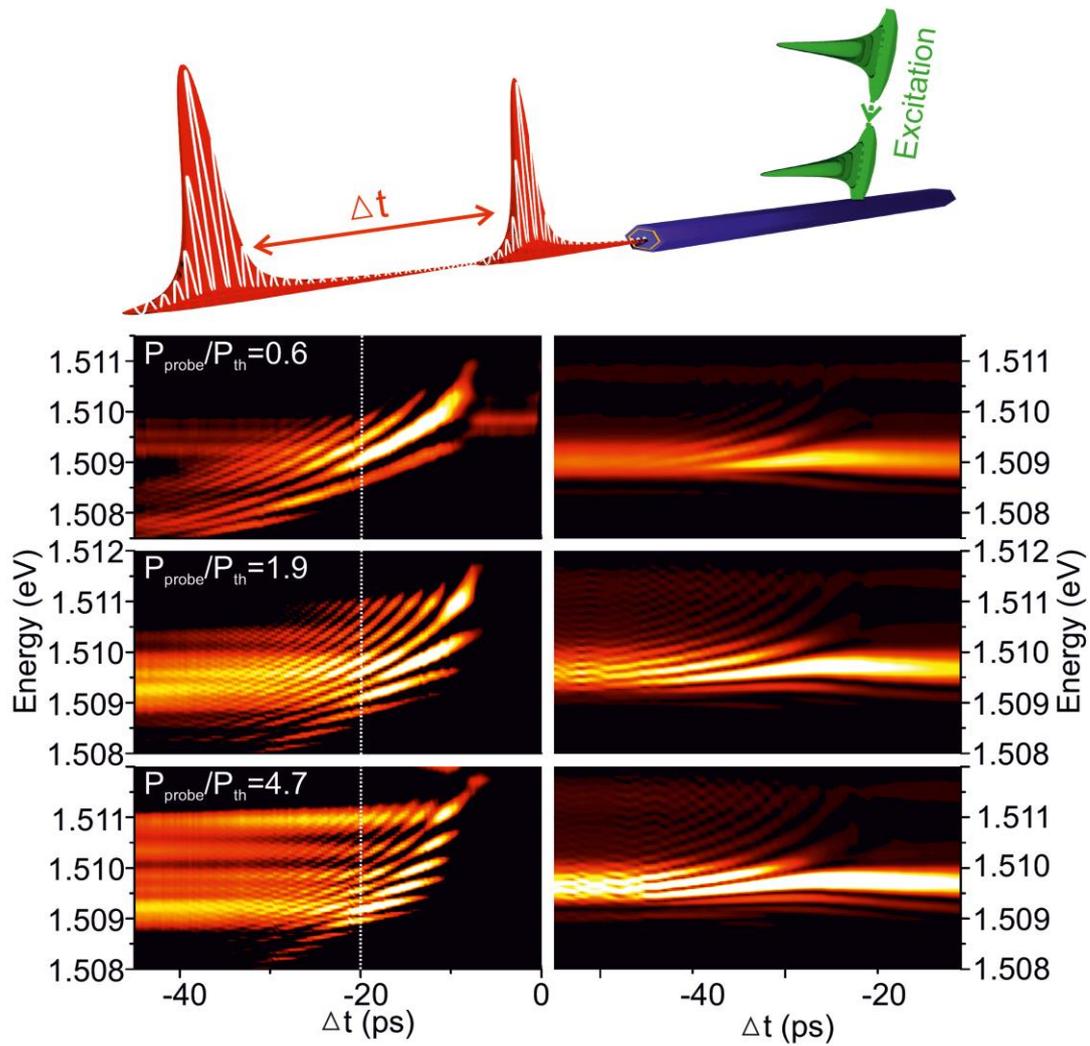

**Figure 1 Time resolved pump-probe response of the NW-laser in experiment (left panels) and simulation (right panel):** NW-laser response to *double* pulse excitation as illustrated schematically in the uppermost figure. The leftmost panesl show pump-probe spectra recorded as a function of $\Delta t$ whereas the righthand column shows the corresponding theoretical results obtained using the optical Bloch equation model described in the text. The three rows depict situations for different probe pulse powers in the SE regime (top row), ASE regime (center row) and lasing regime (bottom row).

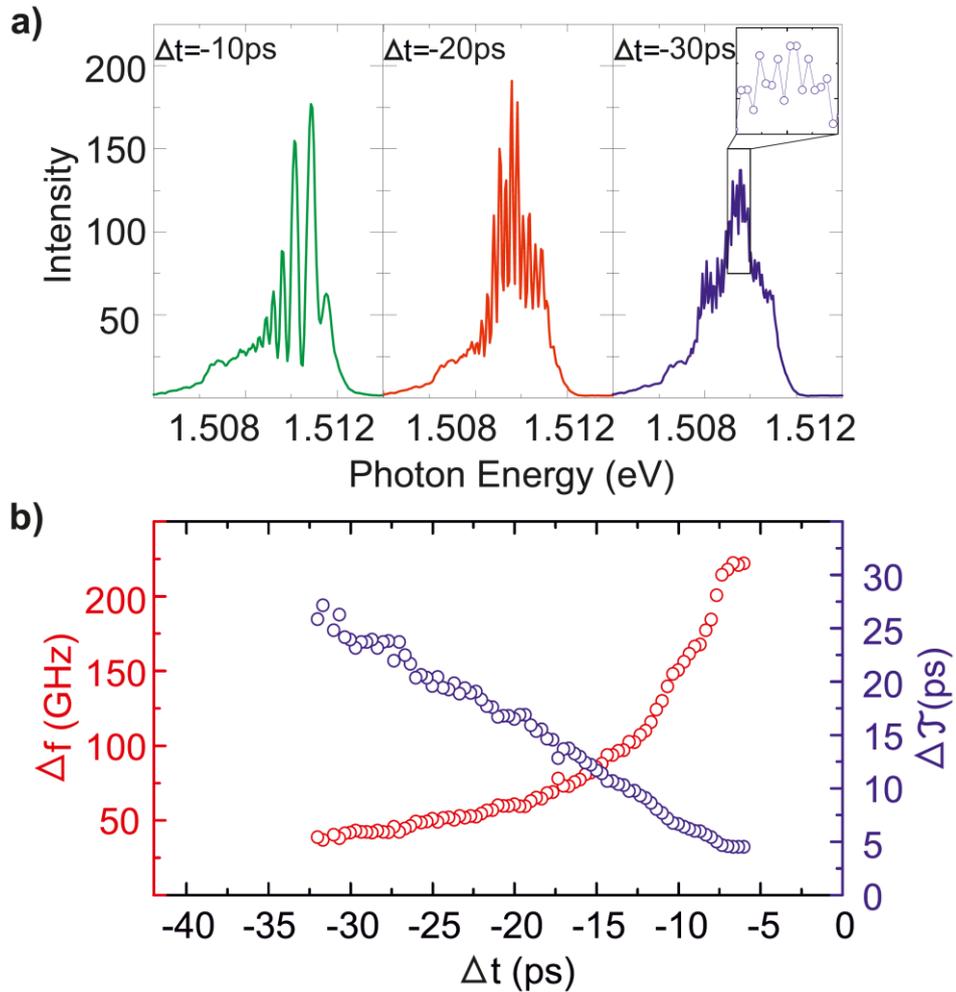

**Figure 2 : Fringe spacing versus pump-probe delay:** (a) Selected spectra recorded from the NW-laser subject to a pump pulse in the lasing regime and a probe pulse in the ASE regime for $\Delta t$=-10ps (green), $\Delta t$=-20ps (red) and $\Delta t$=-30ps (blue). The inset shows a zoom in of the spectrum illustrating that modulation is still observed and that the spacing is close to the resolution limit. (b) repetition rate (red data points) and NW laser pulse separation (blue data points) as a function of $\Delta t$. The data points are measured from the separation of the interference fringes in the NW laser spectra (blue data points) and their inverse (red data points) subject to a pump pulse in the lasing regime and a probe pulse in the ASE regime (example spectra shown in a).

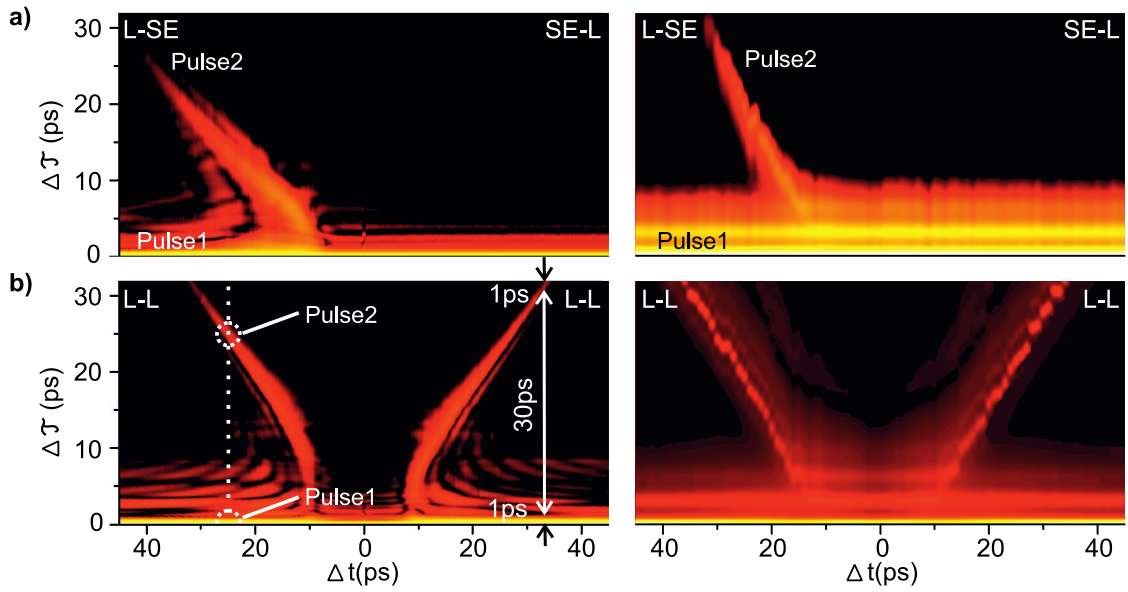

**Figure 3 Experimental (left panels) and simulated (right panels) pump-probe data in the time domain:** Experimental data (left panels) and theoretical calculations (right panels) of the time dependent emission of the NW Laser as a function of $\Delta t$. (a) shows the situation with the pump pulse in the lasing regime and the second (probe) pulse in the SE regime. (b) depicts a situation with both, pump and probe pulse, being in the lasing regime (L-L).

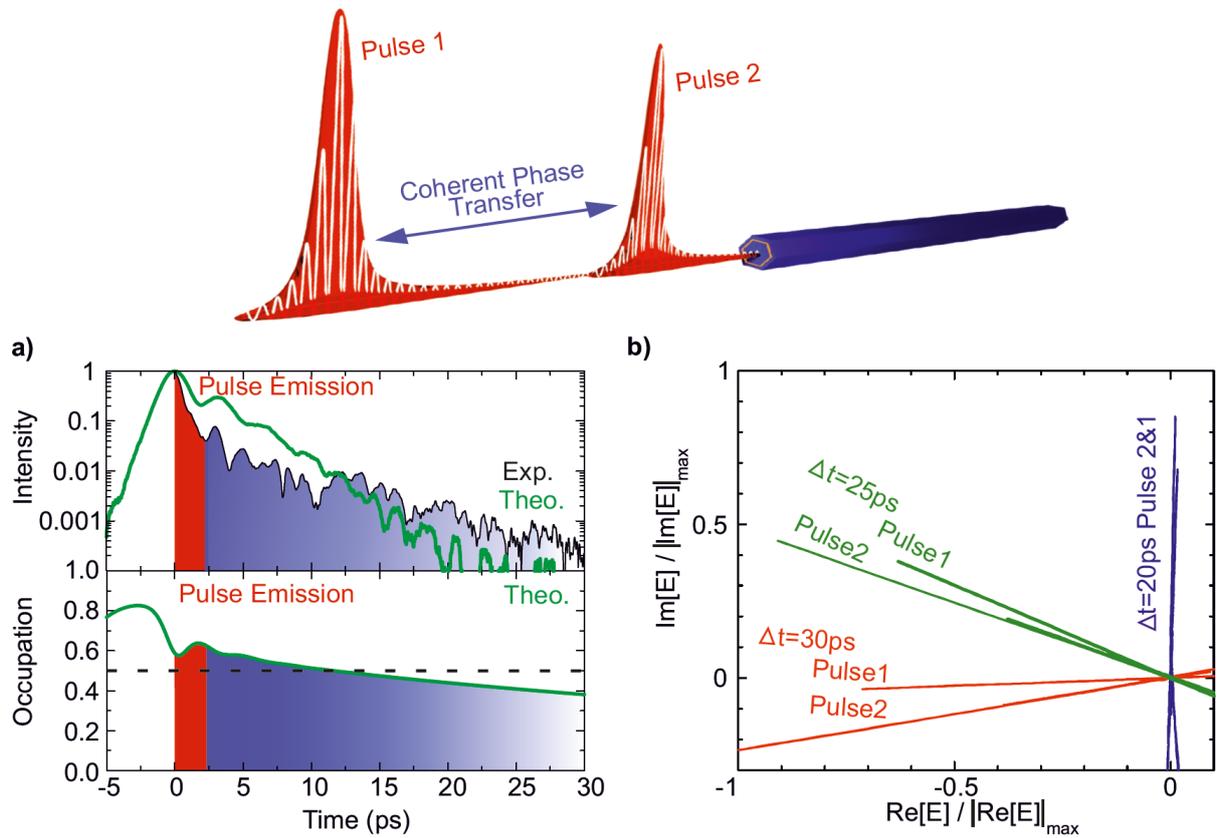

**Figure 4 Coherent phase information transfer between two subsequent NW laser pulses as schematically depicted at the top of the figure:** (a) Time dependence of the emitted intensity in the NW laser (upper panel) and the computed occupation of the lasing state (lower panel) as a function of time. The green curves depict theoretical calculations and the black curve is obtained by dFT of a NW laser peak subject to an excitation pulse with $P/P_{th} = 4$. The red regions mark the initial pulse emission and the blue regions depict Rabi oscillations and coherent phase information transfer. (b) Computed electric field in the complex plane of two subsequent NW laser pulses for $\Delta t =$20ps (blue lines), $\Delta t = 25ps$ (green lines) and $\Delta t = 30ps$ (red lines) illustrating their mutual phase relationship. The values are normalized.

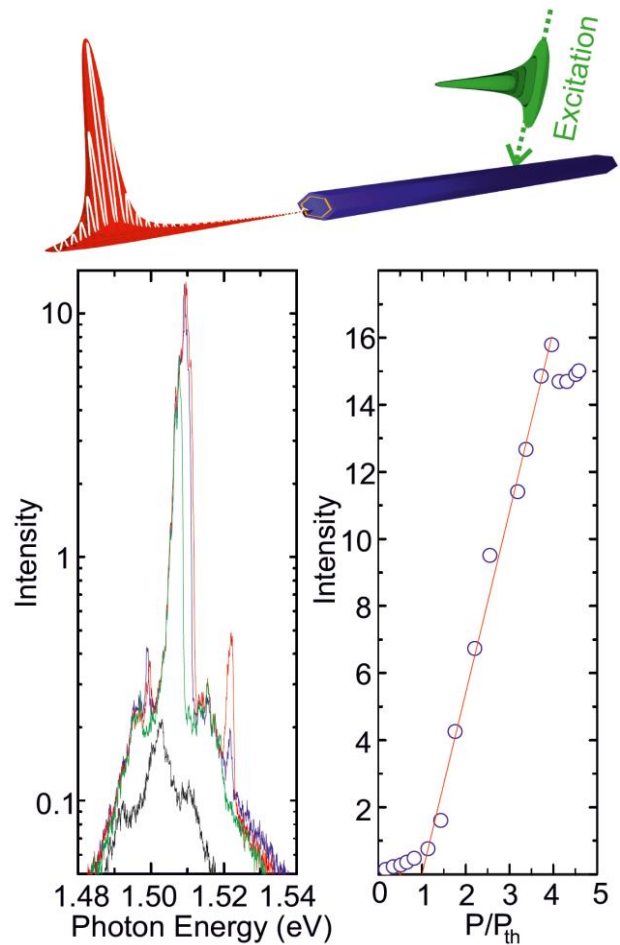

**Figure S1 NW-laser output when excited using a single non-resonant pump pulse as depicted schematically at the top of the figure.** The spectra in the leftmost panel shows selected spectra recorded with P/Pth=0.6 (black curve), P/Pth=1.9 (green curve), P/Pth=4.7 (blue curve) and P/Pth=5.3 (red curve). The rightmost panel shows the characteristic Light-in/Light-out curve obtained from the NW-laser together with the linear fit (red curve) used to determine the lasing threshold.

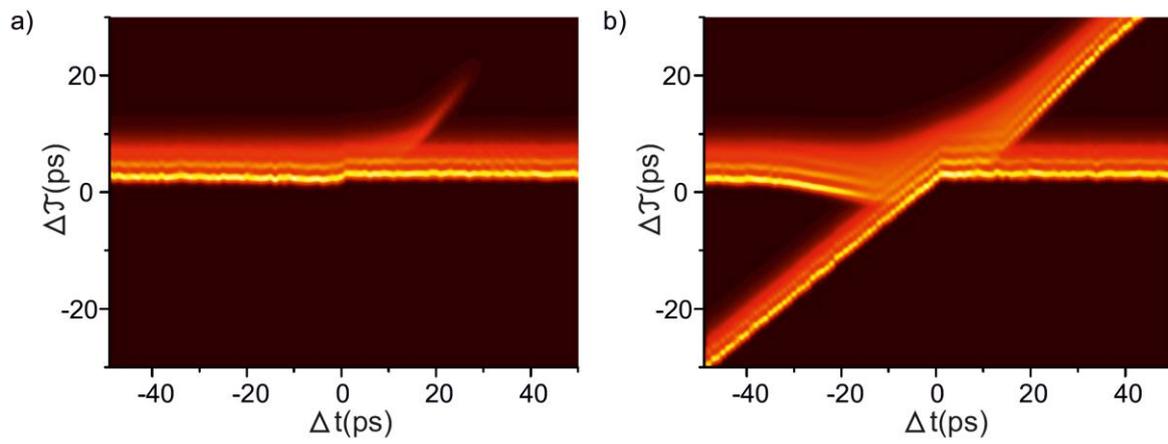

**Figure S2 Theoretical calculations of the time dependent emission of the NW Laser as a function of Δ𝑡.** (a) shows the situation with the pump pulse in the lasing regime and the second (probe) pulse in the SE regime. (b) depicts a situation with both, pump and probe pulse, being in the lasing regime (L-L).